\documentclass[english,aps,prd,preprint,superscriptaddress]{revtex4-1}
\usepackage[latin9]{inputenc}
\usepackage{babel}
\usepackage{mathtools}
\usepackage{amsmath}
\usepackage{amssymb}
\usepackage{graphicx}
\usepackage[unicode=true,pdfusetitle,
 bookmarks=true,bookmarksnumbered=true,bookmarksopen=true,bookmarksopenlevel=1,
 breaklinks=false,pdfborder={0 0 1},backref=false,colorlinks=false]
 {hyperref}

\makeatletter

\usepackage{babel}

\usepackage{babel}

\makeatother

\begin{document}
\title{Viability of Bouncing Cosmology in Energy-Momentum-Squared Gravity}
\author{Ahmed H. Barbar}
\email{ahemdan@aucegypt.edu}

\affiliation{Department of Physics, School of Sciences and Engineering, The American
University in Cairo, AUC Avenue, P.O. Box 74, New Cairo 11835, Egypt}
\author{Adel M. Awad}
\email{awad.adel@aucegypt.edu}

\affiliation{Department of Physics, School of Sciences and Engineering, The American
University in Cairo, AUC Avenue, P.O. Box 74, New Cairo 11835, Egypt}
\affiliation{Department of Physics, Faculty of Science, Ain Shams University, Cairo
11566, Egypt }
\author{Mohammad T. AlFiky}
\email{alfiky@aucegypt.edu}

\affiliation{Department of Physics, School of Sciences and Engineering, The American
University in Cairo, AUC Avenue, P.O. Box 74, New Cairo 11835, Egypt}
\begin{abstract}
We analyze the early-time isotropic cosmology in the so-called energy-momentum-squared
gravity (EMSG). In this theory, a $T_{\mu\nu}T^{\mu\nu}$ term is
added to the Einstein-Hilbert action, which has been shown to replace
the initial singularity by a regular bounce. We show that this is
not the case, and the bouncing solution obtained does not describe
our Universe since it belongs to a different solution branch. The
solution branch that corresponds to our Universe, while nonsingular,
is geodesically incomplete. We analyze the conditions for having viable
regular-bouncing solutions in a general class of theories that modify
gravity by adding higher order matter terms. Applying these conditions
on generalizations of EMSG that add a $\left(T_{\mu\nu}T^{\mu\nu}\right)^{n}$
term to the action, we show that the case of $n=5/8$ is the only
one that can give a viable bouncing solution, while the $n>5/8$ cases
suffer from the same problem as EMSG, i.e. they give nonsingular,
geodesically incomplete solutions. Furthermore, we show that the $1/2<n<5/8$
cases can provide a nonsingular initially de Sitter solution. Finally,
the expanding, geodesically incomplete branch of EMSG or its generalizations
can be combined with its contracting counterpart using junction conditions
to provide a (weakly) singular bouncing solution. We outline the junction
conditions needed for this extension and provide the extended solution
explicitly for EMSG. In this sense, EMSG replaces the standard early-time
singularity by a singular bounce instead of a regular one. 
\end{abstract}
\maketitle

\section{Introduction}

Recent cosmological observations provided us with strong evidence
for the accelerating expansion of the Universe \cite{Riess1998_accelerating,Planck2013}.
Trying to understand this acceleration in general relativity (GR)
one is led to two possibilities: exotic matter content or a cosmological
constant ($\Lambda$). Although these possibilities can fit the observational
data, they do not provide us with a fundamental explanation of this
acceleration. In addition to this large scale problem, GR predicts
its own doom at small scales through the occurrence of spacetime singularities
\cite{hawking1970singularities}, which are expected to be cured in
a full theory of quantum gravity (or at least an effective approximation
of it). These issues have led to a plethora of modified-gravity theories
(see \cite{Clifton2012_mograv_review,Noijri2017_review_modifiedgravity}
for a review, and also \cite{ishak2019testing} for a review on the
recent observational constraints). In these theories, GR is seen as
an effective field theory (of a more general gravitational theory)
that might get corrections either at very large or very small scales.

Some of these modified theories have been shown to replace the initial
cosmological singularity that occurs in GR by a regular bounce (see
\cite{Novello2008_bounce_review,Battefeld2015_bounce_review} for
a review). Along these efforts, energy-momentum-squared gravity (EMSG),
as dubbed by its original authors, was proposed in \cite{EMSG2016}.
This theory modifies gravity by adding a $T_{\mu\nu}T^{\mu\nu}$ term
to the Einstein-Hilbert Lagrangian; it is a special case of theories
that have a Lagrangian of the form $f(R,T_{\mu\nu}T^{\mu\nu})$ which
were first studied in \cite{Nihan2014}.

Further efforts were conducted to study generalizations of EMSG and
their implications. Various cosmological models of higher order generalizations
of EMSG, which add terms in the form $\left(T_{\mu\nu}T^{\mu\nu}\right)^{n}$,
were considered in \cite{board2017cosmological} particularly for
the case $n>1/2$ relevant to high density scales, while the case
$n<1/2$ relevant to late time cosmology was studied in \cite{EMPG2018}.
The cosmological implications for the case $n=1/2$ were studied in
\cite{implications_EMSG2018}, which is interesting since the coupling
in this case becomes dimensionless. Besides the form $\left(T_{\mu\nu}T^{\mu\nu}\right)^{n}$,
a logarithmic generalization, dubbed as energy-momentum log gravity
(EMLG), was considered in \cite{screening2019} where the term $\ln{(\lambda T_{\mu\nu}T^{\mu\nu})}$
was used to extend the $\Lambda$CDM model to study viable cosmologies
and to address the tension in $H_{0}$ measurements. Furthermore,
phenomenological investigations were done in \cite{NeutronStarstConstraintEMSG2018}
using observational data from neutron stars to constrain the free
parameter in EMSG, while in \cite{lowredshift2019} low redshift data
were used to constrain $\left(T_{\mu\nu}T^{\mu\nu}\right)^{n}$ theories.
In addition to these studies, linear stability analysis was used in
\cite{dynamical_in_EMSG2019} to investigate two models in $f(R,T_{\mu\nu}T^{\mu\nu})$
.

In this study, we focus on the analysis of the early-time cosmological
behavior of EMSG and its generalizations, particularly regarding the
existence of regular bounces in this class of theories. As a result,
we show that the bounce obtained in \cite{EMSG2016} is not viable,
and that generic theories that modify GR by adding higher order matter
terms cannot provide a viable regular bounce.

The structure of this paper is as follows. In Sec. \ref{sec:Energy-Momentum-Squared-Gravity},
we review the EMSG theory and its field equations. In Sec. \ref{sec:Cosmology-in-EMSG},
we analyze the isotropic early-time cosmology of EMSG showing that
the bounce obtained in \cite{EMSG2016} is not viable. We show that
the correct solution-branch corresponding to our Universe is also
nonsingular but is not valid beyond a certain point in time, i.e.
past-geodesically incomplete. In Sec. \ref{sec:Bounce-Analysis-for General theories},
we analyze the conditions for having a viable bounce in theories that
modify gravity by adding higher order matter terms. We apply these
conditions to $\left(T_{\mu\nu}T^{\mu\nu}\right)^{n}$ generalizations
of EMSG. In Sec. \ref{sec:Junction-Conditions}, we outline the junction
conditions needed for extending the geodesically incomplete solutions
of EMSG and similar theories. Finally, we conclude with summary and
discussion of the results in Sec. \ref{sec:Conclusion}.

\section{Energy-Momentum-Squared Gravity\label{sec:Energy-Momentum-Squared-Gravity}}

The EMSG action can be written as 
\begin{equation}
S_{\text{EMSG}}=\int d^{4}x\,\sqrt{-g}\,\left(\frac{1}{2\kappa}R-\frac{\Lambda}{\kappa}-\frac{1}{2}\alpha\,T_{\mu\nu}T^{\mu\nu}\right)+S_{\text{M}}\,,
\end{equation}
where $\kappa=8\pi G$, $R$ is the Ricci scalar, $\Lambda$ is the
cosmological constant and $S_{\text{M}}\equiv\int d^{4}x\,\sqrt{-g}\,L_{m}$
with $L_{m}$ as the matter Lagrangian density. Here and thereafter,
we use units where $c=1$ and the metric signature $(-,+,+,+)$.

The extra term that makes this theory different from GR is $-\frac{1}{2}\alpha\,T_{\mu\nu}T^{\mu\nu}$,
where $\alpha$ is a free parameter in the theory (with dimensions
of inverse energy-density) which can be constrained from observations
as was done in \cite{NeutronStarstConstraintEMSG2018}, and $T_{\mu\nu}$
is the ordinary energy-momentum tensor defined as 
\begin{equation}
T_{\mu\nu}=-\frac{2}{\sqrt{-g}}\frac{\delta\left(\sqrt{-g}L_{m}\right)}{\delta g^{\mu\nu}}.
\end{equation}
The factor of $-\frac{1}{2}$, while it can be absorbed into the definition
of $\alpha$ \footnote{$\alpha$ was defined as the free parameter in this way in \cite{NeutronStarstConstraintEMSG2018,EMPG2018}.},
is retained here for convenience. In this case $\alpha$ can be matched
with $\frac{\eta}{\kappa}$ in the original paper \cite{EMSG2016}
where $\eta$ was the free parameter in that case. While $\alpha$
could be positive or negative, we will restrict ourselves to the $\alpha>0$
case since it has been shown to give a more interesting behavior in
early-time cosmology \cite{EMSG2016}.

Now that we have introduced the action of EMSG, let us turn to the
question of how this action can be defined in the first place; that
is, how does the total action already contain the energy-momentum
tensor which is defined by varying part of the total action (i.e.
$S_{\text{M}}$)? The main argument in \cite{EMSG2016,nari2018compact}
is that one does not have to know anything about the gravitational
theory beforehand in order to define the energy-momentum tensor, one
only needs matter physical variables, or simply: the matter Lagrangian
density $L_{m}$. In other words, there is a well-defined way to construct
$T_{\mu\nu}$ in a given theory without gravity, and the $T_{\mu\nu}T^{\mu\nu}$
term (which is just a scalar function in the fields, their derivatives
and the metric) is added as a form of nonminimal coupling of that
theory to gravity.

EMSG is characterized by a density scale $\alpha^{-1}$; deviations
from GR should start appearing near that scale. Since the theory has
a characteristic density scale rather than an energy scale, one can
construct a length scale for physical solutions that have a specific
energy scale. For example, for a charged black hole with a charge
$q$, a characteristic length scale would be $\ell\sim\left(\alpha q^{2}\right)^{1/4}$,
which is the length scale at which the electromagnetic energy density
would be comparable to $\alpha^{-1}$. The physical relevance of that
length can be solution dependent, but the interesting part is that
if the dynamics of the theory impose a maximum density $\sim\alpha^{-1}$,
we get a minimum length $\sim\ell$.

Let us now turn to the field equations. EMSG is equivalent to GR coupled
to an effective matter Lagrangian; therefore the field equations are
just the Einstein equations but sourced by an effective energy-momentum
tensor, and so we have 
\begin{equation}
G_{\mu\nu}+\Lambda g_{\mu\nu}=\kappa T_{\mu\nu}^{\text{eff}},
\end{equation}
where 
\begin{equation}
T_{\mu\nu}^{\text{eff}}=T_{\mu\nu}-\frac{1}{2}\alpha T_{\sigma\rho}T^{\sigma\rho}g_{\mu\nu}+\alpha\Theta_{\mu\nu},
\end{equation}
and 
\begin{equation}
\Theta_{\mu\nu}\equiv\frac{\delta\left(T_{\sigma\rho}T^{\sigma\rho}\right)}{\delta g^{\mu\nu}}=2T_{\mu\sigma}T_{\,\,\nu}^{\sigma}-2L_{m}\left(T_{\mu\nu}-\frac{1}{2}Tg_{\mu\nu}\right)-TT_{\mu\nu}-4T^{\sigma\rho}\frac{\partial^{2}L_{m}}{\partial g^{\mu\nu}\partial g^{\sigma\rho}}.\label{eq:Theta mu nu}
\end{equation}
The details of variation of the extra term can be found in Appendix
\ref{sec:Appendix_Effective-Energy-Momentum-Tensor}.

\section{Cosmology in EMSG\label{sec:Cosmology-in-EMSG}}

Let us start by taking a closer look at the early time cosmology in
EMSG, which was studied in \cite{EMSG2016,NeutronStarstConstraintEMSG2018}.
We will work with a flat Friedmann-Robertson-Walker (FRW) metric $ds^{2}=-dt^{2}+a(t)^{2}\delta_{ij}dx^{i}dx^{j}$.
We will also assume a small positive cosmological constant as in the
usual $\text{\text{\ensuremath{\Lambda}CDM}}$ model. Assuming a perfect
fluid content, we have 
\begin{equation}
T_{\mu\nu}=(\rho+p)\,u_{\mu}u_{\nu}+p\,g_{\mu\nu},
\end{equation}
where $\rho$ is the energy density, $p$ is the pressure and $u^{\mu}$
is the four-velocity of the fluid, which satisfies the conditions
$u^{\mu}u_{\mu}=-1$ and $\nabla_{\nu}\left(u^{\mu}u_{\mu}\right)=0$.
We can arrive at the perfect-fluid energy-momentum tensor through
different Lagrangian densities ($L_{m}=p$ or $L_{m}=-\rho$), which
does not pose a problem in GR \cite{Schutz1970,*Brown_1993_fluid_lagrangian}.
However, in EMSG the Lagrangian density appears explicitly in the
field equations and thus the choice of the Lagrangian density affects
the dynamics. While there is no consensus on which Lagrangian to use
(see \cite{Bertolami2008_nonminimal_fluid_coupling,*Faroni2009_perfect_fluid}
for a detailed discussion), we will stick to the choice of $L_{m}=p$
to follow with the EMSG literature.

For a perfect fluid with $L_{m}=p$, the effective energy momentum
tensor sourcing gravity can be written as 
\begin{equation}
T_{\mu\nu}^{\text{eff}}=(\rho_{\text{eff}}+p_{\text{eff}})\,u_{\mu}u_{\nu}+p_{\text{eff}}\,g_{\mu\nu}.
\end{equation}
where 
\begin{equation}
\rho_{\text{eff}}=\rho-\frac{1}{2}\alpha\left(3p^{2}+8p\rho+\rho^{2}\right),\label{eq:rhoeff}
\end{equation}
\begin{equation}
p_{\text{eff}}=p-\frac{1}{2}\alpha\left(3p^{2}+\rho^{2}\right).\label{eq:peff}
\end{equation}
The effective density and pressure can be defined covariantly as 
\begin{align}
\rho_{\text{eff}} & \coloneqq u^{\mu}u^{\nu}T_{\mu\nu}^{\text{eff}},\\
p_{\text{eff}} & \coloneqq\frac{1}{3}\left(g^{\mu\nu}+u^{\mu}u^{\nu}\right)T_{\mu\nu}^{\text{eff}}.
\end{align}

The Friedmann equations are the same as in GR but with the density
and pressure replaced by their effective counterparts, thus we have
\begin{equation}
H^{2}=\frac{\Lambda}{3}+\frac{\kappa}{3}\rho_{\text{eff}},\label{eq:fried1}
\end{equation}
\begin{equation}
\frac{\ddot{a}}{a}=\dot{H}+H^{2}=\frac{\Lambda}{3}-\frac{\kappa}{6}\left(\rho_{\text{eff}}+3p_{\text{eff}}\right).\label{eq:fried2}
\end{equation}
The remaining equation is the fluid conservation/continuity equation,
which comes from $\nabla^{\mu}T_{\mu\nu}^{\text{eff}}=0$, and of
course can be obtained from the above two equations. Again, this is
nothing more than the fluid conservation equation in GR with the effective
density and pressure, thus we have 
\begin{equation}
\dot{\rho}_{\text{eff}}+3\frac{\dot{a}}{a}\left(\rho_{\text{eff}}+p_{\text{eff}}\right)=0,\label{eq: conseq H in}
\end{equation}
which can be cast into an autonomous form as 
\begin{equation}
\rho_{\text{eff}}^{'}+3\left(\rho_{\text{eff}}+p_{\text{eff}}\right)=0,\label{eq:conseq}
\end{equation}
where the prime denotes derivative with respect to $\ln a$ .

\subsection{Two-component fluids}

Before we attempt to solve the equations, it is worth noting that
because the effective density and pressure are nonlinear in the ordinary
density and pressure, dealing with a multicomponent fluid here will
be drastically different than in GR. In particular, the conservation
equation (\ref{eq:conseq}) will lead to one equation for both fluids.
If we have a two-component fluid, with each component having a barotropic
equation of state in the form $p=\omega\rho$, then the conservation
equation (\ref{eq:conseq}) becomes 
\begin{align}
\left[1-\alpha\left(4\omega_{1}+3\omega_{1}\omega_{2}+4\omega_{2}+1\right)\rho_{2}-\alpha\left(3\omega_{1}^{2}+8\omega_{1}+1\right)\rho_{1}\right]\rho_{1}'\nonumber \\
+\left[1-\alpha\left(4\omega_{1}+3\omega_{1}\omega_{2}+4\omega_{2}+1\right)\rho_{1}-\alpha\left(3\omega_{2}^{2}+8\omega_{2}+1\right)\rho_{2}\right]\rho_{2}'\nonumber \\
+3\left(\omega_{1}+1\right)\rho_{1}-3\alpha\left(3\omega_{1}^{2}+4\omega_{1}+1\right)\rho_{1}^{2}\nonumber \\
+3\left(\omega_{2}+1\right)\rho_{2}-3\alpha\left(3\omega_{2}^{2}+4\omega_{2}+1\right)\rho_{2}^{2}\nonumber \\
-6\alpha\left(2\omega_{1}+3\omega_{1}\omega_{2}+2\omega_{2}+1\right)\rho_{1}\rho_{2}\,\,\, & =0.\label{eq: total conservation equation two fluids}
\end{align}
This is one equation for both components, but we should have individual
equations of motion for each component. Just like for cases other
than perfect fluids, each field has its own equation of motion, and
the conservation of the total energy momentum tensor is satisfied
automatically on shell, i.e. it gives a sum of terms where each term
vanishes on its own when the respective equation of motion is satisfied.
The problem with fluids is that we do not get the continuity or the
Euler equations for each fluid component directly through variation
of the action \footnote{When varying the action of a perfect fluid, one gets a set of equations
that imply particle number conservation and the conservation of the
energy-momentum tensor; see \cite{Brown_1993_fluid_lagrangian} for
a detailed analysis.}; instead, we get those equations for the two fluids combined from
the conservation of the total energy-momentum tensor. Thus, for a
two-component fluid, we should expect to be able to split the conservation
equation into two equations: one equation for each component. In GR,
the split is straightforward since the conservation equation there
is linear; in our case, the splitting seems rather to be an ambiguous
task. We can break this ambiguity using the fact that the Lagrangian
is invariant under exchange of component labels, i.e. $\rho_{1}\longleftrightarrow\rho_{2}$
and $p_{1}\longleftrightarrow p_{2}$. Given this symmetry, we should
expect that the individual-component equations are mapped to one another
under the exchange of labels. Applying this argument on (\ref{eq: total conservation equation two fluids}),
the individual equations for components 1 and 2 respectively are 
\begin{align}
\left[1-\alpha\left(4\omega_{1}+3\omega_{1}\omega_{2}+4\omega_{2}+1\right)\rho_{2}-\alpha\left(3\omega_{1}^{2}+8\omega_{1}+1\right)\rho_{1}\right]\rho_{1}'\nonumber \\
+3\left(\omega_{1}+1\right)\rho_{1}-3\alpha\left(3\omega_{1}^{2}+4\omega_{1}+1\right)\rho_{1}^{2}\nonumber \\
-3\alpha\left(2\omega_{1}+3\omega_{1}\omega_{2}+2\omega_{2}+1\right)\rho_{1}\rho_{2}\,\,\, & =0,\label{eq: cons eq fluid 1}
\end{align}
\begin{align}
\left[1-\alpha\left(4\omega_{1}+3\omega_{1}\omega_{2}+4\omega_{2}+1\right)\rho_{1}-\alpha\left(3\omega_{2}^{2}+8\omega_{2}+1\right)\rho_{2}\right]\rho_{2}'\nonumber \\
+3\left(\omega_{2}+1\right)\rho_{2}-3\alpha\left(3\omega_{2}^{2}+4\omega_{2}+1\right)\rho_{2}^{2}\nonumber \\
-3\alpha\left(2\omega_{1}+3\omega_{1}\omega_{2}+2\omega_{2}+1\right)\rho_{1}\rho_{2}\,\,\, & =0,\label{eq: cons eq fluid 2}
\end{align}
which are interchanged under $\rho_{1}\longleftrightarrow\rho_{2}$
and $\omega_{1}\longleftrightarrow\omega_{2}$. This result can be
easily generalized to the case of an $n$-component fluid by noting
that the Lagrangian then would be invariant under the interchange
of the labels of each pair of components, and the fact that the interaction
terms will still be quadratic in the densities. Thus, the equation
for the $i$th component in that case would be 
\begin{align}
\left(1-\alpha\left(3\omega_{i}^{2}+8\omega_{i}+1\right)\rho_{i}\right)\rho_{i}'+3\left(\omega_{i}+1\right)\rho_{i}-3\alpha\left(3\omega_{i}^{2}+4\omega_{i}+1\right)\rho_{i}^{2}\nonumber \\
-\alpha\sum_{j\neq i}^{n}\left\{ \left(4\omega_{i}+3\omega_{i}\omega_{j}+4\omega_{j}+1\right)\rho_{j}\rho_{i}'\right.\nonumber \\
\left.+3\left(2\omega_{i}+3\omega_{i}\omega_{j}+2\omega_{j}+1\right)\rho_{i}\rho_{j}\right\} \,\, & =0.
\end{align}

Now that we have an equation of motion for each fluid component, it
is important to note that these equations are the ones that determine
how the fluid density of each component behaves with the scale $a$.
For example, in GR the solutions for matter and radiation are $\rho_{m}\sim a^{-3}$
and $\rho_{r}\sim a^{-4}$, which tells us that radiation is the dominant
component in the early Universe where $a$ is very small. In our case,
we need to solve Eqs. (\ref{eq: cons eq fluid 1}) and (\ref{eq: cons eq fluid 2})
simultaneously for matter and radiation content, which can be quite
difficult analytically without resorting to some kind of approximation.
In Appendix \ref{sec:Appendix_Two-component-Fluid-in}, we show that
indeed radiation will be dominant in the early Universe in EMSG.

\subsection{Radiation domination}

Before focusing on radiation, let us start by considering a general
one-component fluid in EMSG. In what follows, although $\Lambda$
can be ignored since we are interested only in the early Universe,
we will keep it for reasons to be clear shortly.

For a one-component fluid with a barotropic equation of state of the
form $p=\omega\rho$, the effective density and pressure in (\ref{eq:rhoeff})
and (\ref{eq:peff}) become 
\begin{equation}
\rho_{\text{eff}}=\rho-\frac{1}{2}\left(3\omega^{2}+8\omega+1\right)\alpha\rho^{2},\label{eq:rho_eff general omega}
\end{equation}
\begin{equation}
p_{\text{eff}}=\omega\rho-\frac{1}{2}\left(3\omega^{2}+1\right)\alpha\rho^{2}.\label{eq:p_eff general omega}
\end{equation}
At this point it is worth noting that one can find values for $\omega$
such that the effective equation of state maintains the same form
as the original equation of state, i.e. $p_{\text{eff}}=\omega\rho_{\text{eff}}.$
It is easy to see from (\ref{eq:rho_eff general omega}) and (\ref{eq:p_eff general omega})
that these special values for $\omega$ satisfy the following cubic
equation: 
\begin{equation}
3\omega^{2}+1=\omega\left(3\omega^{2}+8\omega+1\right).
\end{equation}
This gives the solutions $\omega=-1$ and $\omega=1/3$ ($\omega=-1$
is a repeated root). Given this result, it would be interesting to
consider a dark energy ($\omega=-1$) fluid in EMSG, but in this paper
we will concern ourselves with a $\text{\text{\ensuremath{\Lambda}CDM}}$-like
model in EMSG with nonexotic fluids (i.e. only matter and radiation).
We note that these specific values for $\omega$ that preserve the
equation of state are nothing but a coincidence in EMSG, and other
values (if any) would appear in higher order generalizations.

In general, $p_{\text{eff}}$ is not always single valued in $\rho_{\text{eff}}$;
this is due to the fact that (\ref{eq:rho_eff general omega}) is
not invertible over the entire domain of $\rho$; however, we can
put it in the form $p_{\text{eff}}=\omega_{\text{eff}}(\rho)\,\rho_{\text{eff}}$
by dividing (\ref{eq:p_eff general omega}) by (\ref{eq:rho_eff general omega})
to get $\omega_{\text{eff}}$ as 
\begin{align}
\omega_{\text{eff}}(\rho) & =\frac{\omega-\frac{1}{2}\left(3\omega^{2}+1\right)\alpha\rho}{1-\frac{1}{2}\left(3\omega^{2}+8\omega+1\right)\alpha\rho}.
\end{align}

As we can see from (\ref{eq:rho_eff general omega}), for values of
$\omega$ that keep $3\omega^{2}+8\omega+1>0$, $\rho_{\text{eff}}$
is not a monotonically increasing function of $\rho$ and can be zero
(or even be negative) for $\rho>0$; this feature is what makes this
theory appealing, since it opens the possibility of having a critical
point $H=0$ at high densities {[}which can be seen directly from
(\ref{eq:fried1}){]}, similar to other theories, like loop quantum
gravity \cite{Ashtekar2006_BouncingLoop,*Bounce_in_LQC_2006} or braneworlds
\cite{shtanov2003bouncing_bw} that have a Friedmann equation of the
form 
\begin{equation}
H^{2}\sim\rho(1-\frac{\rho}{\rho_{\text{critical}}}),
\end{equation}
where in EMSG $\rho_{\text{critical}}$ would be $O(\alpha^{-1})$.

If we look for a critical point $H=0$ in EMSG, we will need {[}from
(\ref{eq:fried1}){]} to have $\rho_{\text{eff}}=-\frac{\Lambda}{\kappa}$,
which from (\ref{eq:rho_eff general omega}) gives the critical density
as 
\begin{equation}
\rho_{\text{critical}}=\frac{1+\sqrt{1+2\alpha\frac{\Lambda}{\kappa}\left(3\omega^{2}+8\omega+1\right)}}{\alpha\left(3\omega^{2}+8\omega+1\right)},\label{eq:rho critical general}
\end{equation}
where we have discarded the other solution as it leads to $\rho\simeq-\frac{\Lambda}{\kappa}+O\left(\frac{\alpha\Lambda^{2}}{\kappa^{2}}\right)$.
At the critical density, using the expressions for $\rho_{\text{eff}}$
and $p_{\text{eff}}$ in (\ref{eq:rho_eff general omega}) and (\ref{eq:p_eff general omega}),
the acceleration equation (\ref{eq:fried2}) becomes 
\begin{equation}
\frac{\ddot{a}}{a}=\dot{H}=\frac{(\omega+1)(3\omega+1)}{(3\omega^{2}+8\omega+1)}\Lambda-\frac{\kappa(\omega+1)^{2}(3\omega-1)}{2\alpha(3\omega^{2}+8\omega+1)^{2}}\left(1+\sqrt{1+2\alpha\frac{\Lambda}{\kappa}\left(3\omega^{2}+8\omega+1\right)}\right).\label{eq: fred2 at critical point}
\end{equation}
It is easy to check, given that we expect $\alpha^{-1}\ggg\frac{\Lambda}{\kappa}$
, that $\dot{H}<0$ for $\omega>1/3$ as the second term in (\ref{eq: fred2 at critical point})
dominates and it is negative for those values of $\omega$. Therefore,
the critical point we found corresponds to a bounce only for $\omega\leq1/3$.

\subsubsection{Bounce point and its viability}

Let us now turn our attention to radiation ($\omega=1/3$), which
has an effective density and pressure as 
\begin{align}
\rho_{\text{eff}} & =\rho_{r}-2\alpha\rho_{r}^{2},\label{eq:rho eff for radiation}\\
p_{\text{eff}} & =\frac{\rho_{r}}{3}-\frac{2\alpha\rho_{r}^{2}}{3}.\label{eq:peff for radiation}
\end{align}
At the critical point, from (\ref{eq:rho critical general}), the
radiation density will be 
\begin{equation}
\rho_{r\,\text{crit.}}=\frac{1}{4\alpha}\left(1+\sqrt{1+8\alpha\frac{\Lambda}{\kappa}}\right),\label{eq:rho rad at H=00003D00003D0}
\end{equation}
and as we see from (\ref{eq: fred2 at critical point}), this gives
$\dot{H}=\frac{2\Lambda}{3}$. This was the main reason of keeping
$\Lambda$ explicit up until this point, to show that we can get $\dot{H}>0$
at the critical point due to $\Lambda>0$, which was an argument in
\cite{EMSG2016} for the necessity of having a positive cosmological
constant in this theory. It is worth noting, however, that if we had
not ignored the matter fluid component, it could have led to a similar
result, i.e. $\dot{H}\sim\rho_{m}>0$ without the need for any cosmological
constant. In either case, one can conclude that the point at which
the density is as in (\ref{eq:rho rad at H=00003D00003D0}) corresponds
to a regular bounce as was concluded in \cite{EMSG2016}. However,
we will show that this critical point does not correspond to a solution
that describes our Universe. To see this, let us solve the conservation
equation explicitly. From (\ref{eq:rho eff for radiation}), (\ref{eq:peff for radiation})
and (\ref{eq:conseq}), we get the conservation equation for radiation
as 
\begin{equation}
\left(1-4\alpha\rho_{r}\right)\rho_{r}'+4\rho_{r}-8\alpha\rho_{r}^{2}=0.\label{eq:conseq radiation only}
\end{equation}
Notice that this ordinary differential equation (ODE) has a regular
singular point at $\rho_{r}=\frac{1}{4\alpha}$, we will discuss the
relevance of this issue later. Since this is an autonomous ODE with
respect to $\ln a$ , it can be integrated directly to give $\ln a$
as a function of $\rho_{r}$ . Integrating with the condition $a(\rho_{r0})=a_{0}$,
where $\rho_{r0}$ is the cosmological radiation density at the present
time \footnote{One might object to the use of $\rho_{r0}$ here, but it is important
to note that, since EMSG has to reduce to GR while still being in
the radiation dominated era, Eq. (\ref{eq:conseq radiation only})
reduces to the standard conservation equation for radiation (which
is effectively decoupled from other fluid components) where $\rho_{r0}$
can be used.}, and setting $a_{0}=1$, we get 
\begin{equation}
a=\left(\frac{\rho_{r0}\left(1-2\alpha\rho_{r0}\right)}{\rho_{r}\left(1-2\alpha\rho_{r}\right)}\right)^{1/4}.\label{eq: scale as a function of rho_r}
\end{equation}
This expression reduces to the usual $a=\left(\frac{\rho_{r0}}{\rho_{r}}\right)^{1/4}$
for low densities (i.e. in the limit $\alpha\rho_{r}\rightarrow0$).
Since $a$ has to be real, we must have 
\begin{equation}
\rho_{r}<\frac{1}{2\alpha}.\label{eq: constraint on rho_r}
\end{equation}
Notice that having $\rho_{r0}<\frac{1}{2\alpha}$ is the main cause
of this constraint. The density at the critical point in (\ref{eq:rho rad at H=00003D00003D0})
clearly violates (\ref{eq: constraint on rho_r}), and thus the critical
point is unphysical and there is no bouncing solution that corresponds
to our Universe. In a hypothetical universe where $\rho_{r0}>\frac{1}{2\alpha}$,
the requirement that $a$ has to be real would have led instead to
$\rho_{r}>\frac{1}{2\alpha}$, and the critical point (\ref{eq:rho rad at H=00003D00003D0})
would have corresponded indeed to a bouncing solution in that universe.
In other words, EMSG gives a valid regular bounce in a universe where
the density is always higher than $\frac{1}{2\alpha}$.

\subsubsection{Radiation domination solution}

Solving for $\rho_{r}$ in (\ref{eq: scale as a function of rho_r}),
we get 
\begin{equation}
\rho_{r}(a)=\frac{1}{4\alpha}\left(1\pm\sqrt{1-8\alpha\rho_{r0}\left(1-2\alpha\rho_{r0}\right)a^{-4}}\right).\label{eq:rho r sol branches}
\end{equation}
We pick the negative branch because it gives the asymptotic behavior
of $\rho_{r}\simeq\rho_{r0}a^{-4}$ for relatively large $a$. Since
we want $\rho_{r}$ to be real, we must have a maximum density $\rho_{r\,\text{max}}=\frac{1}{4\alpha}$
corresponding to a minimum scale factor. We can then write the solution
as 
\begin{equation}
\rho_{r}(a)=\frac{1}{4\alpha}\left(1-\sqrt{1-\left(\frac{a_{\text{min}}}{a}\right)^{4}}\right),\label{eq:rho_r solution}
\end{equation}
where 
\begin{equation}
a_{\text{min}}\equiv\left(8\alpha\rho_{r0}\left(1-2\alpha\rho_{r0}\right)\right)^{1/4}.
\end{equation}
The existence of a minimum scale factor here comes from the constraint
$\rho_{r}\in\mathbb{R}$ rather than from the dynamical solution $a(t)$
as in the case of a bounce; this will lead to geodesic incompletion
as we will see shortly. One can easily get $a(t)$ by plugging (\ref{eq:rho_r solution})
in the Friedmann equation (\ref{eq:fried1}). An equivalent, but clearer,
way is to notice that since $p_{\text{eff}}=\frac{1}{3}\rho_{\text{eff}}$,
the conservation equation (\ref{eq:conseq}) becomes 
\begin{equation}
\rho_{\text{eff}}^{'}+4\rho_{\text{eff}}=0;
\end{equation}
we can solve this with the already known condition that $\rho_{r}(a_{\text{min}})=\frac{1}{4\alpha}$,
which gives us $\rho_{\text{eff}}(a_{\text{min}})=\frac{1}{8\alpha}$.
With the latter condition, we get the solution 
\begin{equation}
\rho_{\text{eff}}(a)=\frac{1}{8\alpha}\left(\frac{a_{\text{min}}}{a}\right)^{4}.\label{eq:rho_eff of a}
\end{equation}
Plugging (\ref{eq:rho_eff of a}) in the Friedmann equation (\ref{eq:fried1}),
we get 
\begin{equation}
H^{2}=\frac{\Lambda}{3}+\frac{\kappa}{24\alpha}\left(\frac{a_{\text{min}}}{a}\right)^{4}.\label{eq:Friedmann 1 with a}
\end{equation}
We can clearly see now that there are no $H=0$ critical points for
all real (physical) values of $a$. From now on, for simplicity, we
shall ignore the cosmological constant in the early Universe; it has
served its purpose now that we have established that the critical
point in (\ref{eq:rho rad at H=00003D00003D0}) does not correspond
to a bounce. The Friedmann equation now becomes 
\begin{equation}
H=\frac{\dot{a}}{a}=\pm\sqrt{\frac{\kappa}{24\alpha}}\left(\frac{a_{\text{min}}}{a}\right)^{2}.\label{eq:a_dot ODE}
\end{equation}
We can conveniently define 
\begin{equation}
H_{\text{max}}\equiv H(a_{\text{min}})=\sqrt{\frac{\kappa}{24\alpha}}.\label{eq:H_max}
\end{equation}
Solving the positive branch of (\ref{eq:a_dot ODE}) with the condition
$a(0)=a_{\text{min}}$, we get 
\begin{equation}
a(t)=a_{\text{min}}\sqrt{1+2H_{\text{max}}t},\,\,\,\,\,\,\,t\geq0.\label{eq: a(t) t>0 1st}
\end{equation}
This solution, which was found in \cite{NeutronStarstConstraintEMSG2018}
{[}albeit with a redefinition of $t$ to match the standard solution
of GR at which $a(0)=0${]}, manifestly cannot be extended for $t<\frac{-1}{2H_{\text{max}}}$,
but more importantly, we cannot extend it for $t<0$ since that would
lead to $a<a_{\text{min}}$ and then from (\ref{eq:rho_r solution})
the radiation density would become nonreal as we discussed before.
This solution can be interpreted, in the spirit of effective field
theory, as EMSG breaking down as we approach $a_{\text{min}}$ and
one would need new physics to describe what is happening beyond that
point. In this sense, $a_{\text{min}}$ is not interpreted as an absolute
minimum scale of nature, but the minimum scale at which EMSG is valid.

\begin{figure}
\centering{}\includegraphics[width=8.6cm]{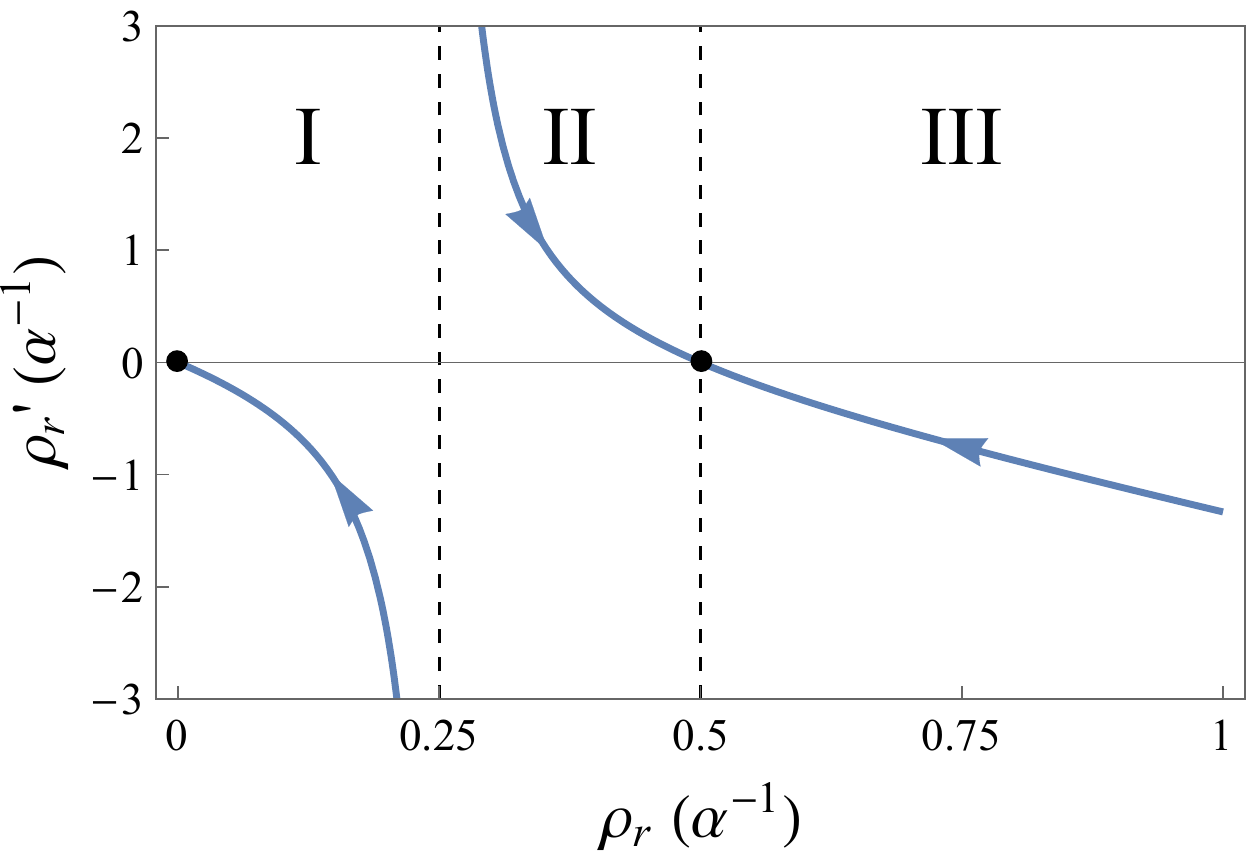}\caption{\label{fig:Phase-Plot-for radiation EMSG}Phase plot for Eq. (\ref{eq:conseq radiation only})
in units of $\alpha^{-1}$. The dots represent fixed points, while
the arrows represent the flow direction in phase space. Region I represents
the negative branch of (\ref{eq:rho r sol branches}), while region
II represents the positive branch. Region III contains the bounce
point, which is disconnected from the physical region that describes
our Universe.}
\end{figure}

It is interesting to note that while $\dot{\rho_{r}}$ diverges as
$t\rightarrow0$, all geometric quantities {[}represented by $H(t)$
and its time derivatives{]} remain finite. This gives us the chance
to extend the spacetime beyond $t=0$ by combining the solution in
(\ref{eq: a(t) t>0 1st}) with its counterpart from the negative branch
of (\ref{eq:a_dot ODE}) using junction conditions at $t=0$. We will
present this in Sec. \ref{sec:Junction-Conditions}.

We conclude this section by reflecting on the issue that prevented
this theory from achieving a cosmologically viable bounce, after all,
it had a Friedmann equation reminiscent of theories like loop quantum
gravity and braneworlds, so why is the case here different? The reason
is that, unlike those theories which modify only the Friedmann equation,
EMSG also modifies the conservation/continuity equation. The main
issue is that, due to the nonlinearity, the equation is modified in
a singular way; particularly, the singular point is at a lower density
($\rho_{r}=\frac{1}{4\alpha}$) than the density at the critical point
(\ref{eq:rho rad at H=00003D00003D0}). This causes our Universe to
be in a solution-region entirely disconnected from the bounce point;
we can see this from the phase plot of the conservation equation (\ref{eq:conseq radiation only})
in Fig. \ref{fig:Phase-Plot-for radiation EMSG}. We note that this
problem is not unique to EMSG; it can happen in any other theory that
effectively modifies the matter Lagrangian. We discuss the conditions
for this issue in the next section.

\section{Bounces in More General Theories\label{sec:Bounce-Analysis-for General theories}}

Let us start with a more generalized theory than EMSG that effectively
modifies the matter Lagrangian but keeps the geometric side as GR,
so we would have a total action like 
\begin{equation}
S=\int d^{4}x\,\sqrt{-g}\,\left(\frac{1}{2\kappa}R+L_{m\,\text{eff}}\right).
\end{equation}
We assume FRW metric and a perfect fluid matter content with a generic
equation of state $p=p(\rho)$. We will focus only on a single-component
fluid or a fluid with one dominant component in the early Universe
(which typically should be the radiation). As in the case with EMSG,
the Friedmann equations take the form 
\begin{equation}
H^{2}=\frac{\kappa}{3}\rho_{\text{eff}},\label{eq:fried1-1}
\end{equation}
\begin{equation}
\dot{H}+H^{2}=-\frac{\kappa}{6}\left(\rho_{\text{eff}}+3p_{\text{eff}}\right).\label{eq:fried2-1}
\end{equation}
It is useful to combine these equations to get $\dot{H}$ in terms
of $\rho_{\text{eff}}$ and $p_{\text{eff}}$ as 
\begin{equation}
\dot{H}=-\frac{\kappa}{2}\left(\rho_{\text{eff}}+p_{\text{eff}}\right).\label{eq: H_dot in terms of eff}
\end{equation}
We have ignored the cosmological constant for simplicity, but the
arguments below can be easily generalized by absorbing the cosmological
constant in the definition of $\rho_{\text{eff}}$ and $p_{\text{eff}}$.
The conservation equation is the same as (\ref{eq: conseq H in}),
which can be written in terms of $\dot{\rho}$ as 
\begin{equation}
\dot{\rho}=-3H\left(\frac{\rho_{\text{eff}}+p_{\text{eff}}}{\frac{d\rho_{\text{eff}}}{d\rho}}\right).\label{eq: rho_dot general}
\end{equation}
We want to analyze the conditions at which this theory would give
a cosmologically viable regular bounce. From (\ref{eq:fried1-1}),
(\ref{eq: H_dot in terms of eff}) and (\ref{eq: rho_dot general}),
we see that the behavior of $H$, $\dot{H}$ and $\dot{\rho}$ is
controlled by only three functions: $\rho_{\text{eff}}$, $\rho_{\text{eff}}+p_{\text{eff}}$
and $\frac{d\rho_{\text{eff}}}{d\rho}$, which are all functions in
$\rho$. So we can take $\rho$ as the basic variable that controls
the phase space of this dynamical system. We assume for simplicity
that these functions do not have more than one nontrivial zero; the
arguments in this section can be generalized otherwise. We also assume
that these functions are continuous and smooth as functions of $\rho$;
this assumption is important in order to avoid curvature-singularity
problems at finite values of $\rho$.

\begin{figure}
\centering{}\includegraphics[width=8.6cm]{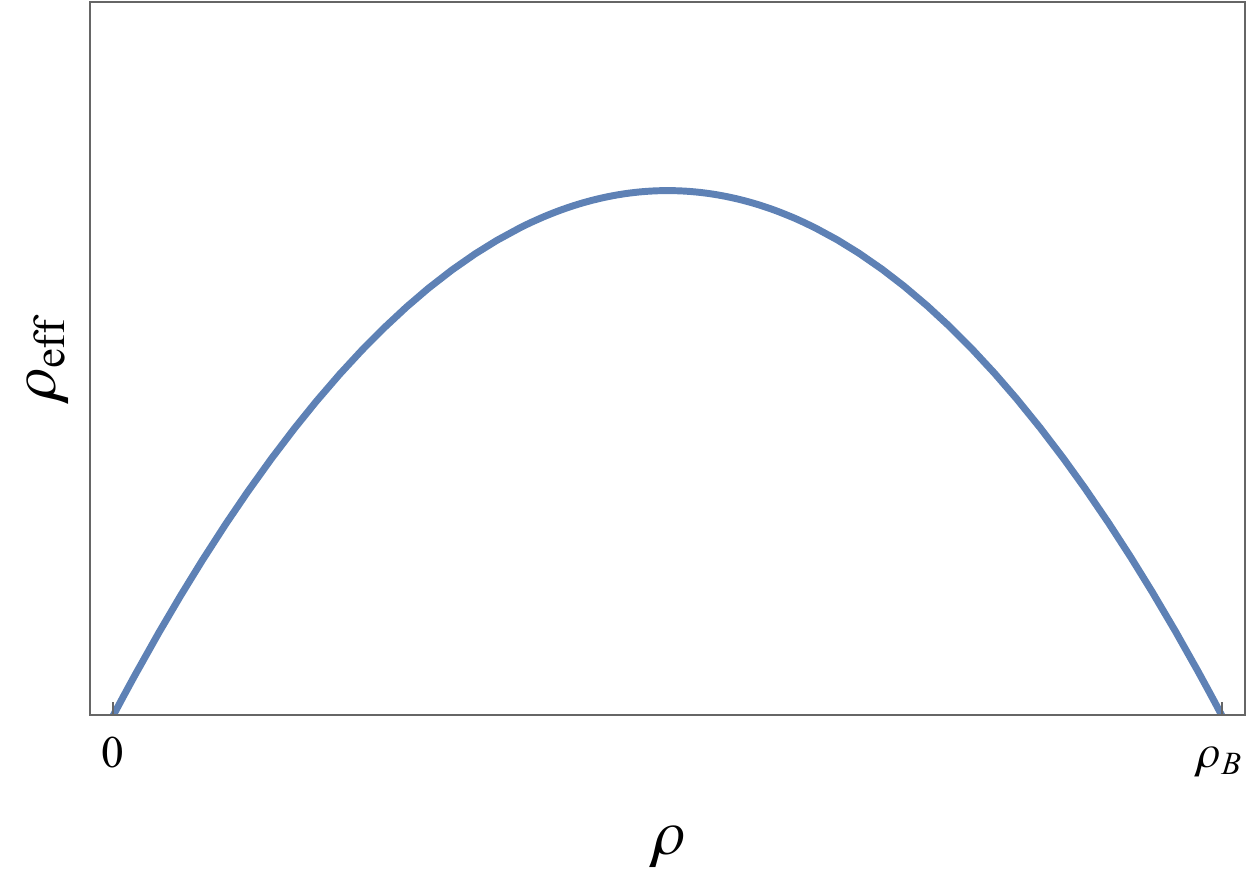}\caption{\label{fig:rhp_eff profile example}An example profile for the effective
density as a function of the ordinary density $\rho$.}
\end{figure}

\subsection{Bounce analysis}

In order for this theory to have a bounce at some high density $\rho_{\text{B}}$,
the usual bounce conditions of GR, $H=0$ and $\dot{H}>0$, must be
satisfied at that point. These conditions then imply the following
from (\ref{eq:fried1-1}) and (\ref{eq: H_dot in terms of eff}):
\begin{equation}
\rho_{\text{eff}}\Bigr|_{\rho=\rho_{\text{B}}}=0,\label{eq: bounce cond 1}
\end{equation}
\begin{equation}
(\rho_{\text{eff}}+p_{\text{eff}})\Bigr|_{\rho=\rho_{\text{B}}}<0.\label{eq: bounce cond 2}
\end{equation}
For cosmological viability, the low density behavior must be the same
as in GR, this implies 
\begin{equation}
\rho_{\text{eff}}\Bigr|_{\rho\ll\rho_{\text{B}}}\simeq\rho>0,\label{eq: low density cond 1}
\end{equation}
\begin{equation}
(\rho_{\text{eff}}+p_{\text{eff}})\Bigr|_{\rho\ll\rho_{\text{B}}}\simeq\rho+p>0.\label{eq: low density cond 2}
\end{equation}
To reconcile the conditions (\ref{eq: bounce cond 1}) and (\ref{eq: bounce cond 2})
with (\ref{eq: low density cond 1}) and (\ref{eq: low density cond 2}),
each of $\rho_{\text{eff}}$ and $\rho_{\text{eff}}+p_{\text{eff}}$
must have at least one local maximum in the interval $\left]\rho_{0},\rho_{\text{B}}\right[$,
where $\rho_{0}\ll\rho_{\text{B}}$ is the density observed at the
present time. For simplicity we assume that $\rho_{\text{eff}}$ and
$\rho_{\text{eff}}+p_{\text{eff}}$ each has only one maximum in that
interval, so that their behavior as functions of $\rho$ is as follows:
they start monotonically increasing, hit a maximum, then they become
monotonically decreasing. A simple profile for these functions is
shown in Fig. \ref{fig:rhp_eff profile example}. This behavior with
(\ref{eq: bounce cond 2}) implies that $\rho_{\text{eff}}+p_{\text{eff}}$
must zero-cross at a point $\rho_{\text{C}}<\rho_{\text{B}}$, or
in other words 
\begin{equation}
\exists\,\rho_{\text{C}}<\rho_{\text{B}}\,\,:\,\,(\rho_{\text{eff}}+p_{\text{eff}})\Bigr|_{\rho=\rho_{\text{C}}}=0.\label{eq: rho_C condition}
\end{equation}
Let the maximum of $\rho_{\text{eff}}$ be denoted by $\rho_{\text{A}}$.
Thus from the discussion above 
\begin{equation}
\exists\,\rho_{\text{A}}<\rho_{\text{B}}\,\,:\,\,\frac{d\rho_{\text{eff}}}{d\rho}\Bigr|_{\rho=\rho_{\text{A}}}=0.\label{eq: rho_A conditon}
\end{equation}
Note that since $\rho_{A}$ is a maximum, it is a zero of $\frac{d\rho_{\text{eff}}}{d\rho}$
with odd multiplicity.

Let us now look at the structure of the phase space of our dynamical
system, which can be described from (\ref{eq: H_dot in terms of eff})
and (\ref{eq: rho_dot general}) by the behavior of $\dot{H}$ and
$\dot{\rho}$ as functions of $\rho$ \footnote{Similar analyses can be done using $H$ as the main phase space variable,
e.g. see \cite{Awad2013_FixedPoints}}. We are interested in fixed and singular points. A fixed point of
the system is a point at which $\dot{\rho}=\dot{H}=0$. If the system
starts at a fixed point it stays there forever (provided that the
system is at least Lipschitz continuous at the fixed point), and if
a system starts at a nonfixed point, it takes an infinite time to
reach a fixed point; the latter fact can be easily deduced from the
time reversal of the former one. We see from (\ref{eq: H_dot in terms of eff})
and (\ref{eq: rho_dot general}) that the fixed points of our system
are only described by the zeros of $\rho_{\text{eff}}+p_{\text{eff}}$
(and thus our system is Lipschitz continuous at fixed points from
our assumptions on $\rho_{\text{eff}}+p_{\text{eff}}$). A singular
point is a point at which either $\dot{\rho}$ or $\dot{H}$ diverges.
$\dot{H}$ is well behaved from our assumption about continuity and
smoothness of $\rho_{\text{eff}}+p_{\text{eff}}$, so we only need
to focus on singular points of $\dot{\rho}$. By using the auxiliary
equations 
\begin{equation}
\dot{\rho}=H\rho',
\end{equation}
\begin{equation}
\rho'=\frac{-3\left(\rho_{\text{eff}}+p_{\text{eff}}\right)}{\frac{d\rho_{\text{eff}}}{d\rho}},\label{eq: rho prime general}
\end{equation}
where $\rho'$ is the first derivative of $\rho$ with respect to
$\ln a$, we see that (\ref{eq: rho prime general}) captures both
the fixed and the singular points of the system; thus, it is sufficient
to turn our focus into the sub-phase-space of $(\rho',\rho)$ for
our analysis of these points.

Before proceeding, we need to show the following statement: 
\begin{quote}
\textit{For an autonomous dynamical system (Lipschitz continuous at
fixed points) controlled by a variable $\rho(t)$, if $\rho_{*}$
is either a fixed point of the system or a singular point with odd
multiplicity, then the phase space is split at $\rho_{*}$ into two
regions: $\rho<\rho_{*}$ and $\rho>\rho_{*}$,} 
\end{quote}
where split means that if the system starts in the region $\rho<\rho_{*}$
it cannot reach---either backward or forward in time---a point in
the region $\rho>\rho_{*}$ ( in a finite time) and vice versa, and
a singular point with odd multiplicity means that $\rho'$ switches
signs after crossing $\rho_{*}$, which can only happen if $\rho'$
has a pole at $\rho_{*}$ with odd multiplicity.

Showing the above statement for a fixed point is very straightforward:
if the system starts in the region $\rho<\rho_{*}$, it takes an infinite
time to reach $\rho_{*}$ let alone cross it and vice versa. In the
case where $\rho_{*}$ is a singular point with odd multiplicity,
it will act either as an attractive (sink) or a repulsive (source)
point in the phase space, which splits it into two regions.

\begin{figure}
\centering{}\includegraphics[width=8.6cm]{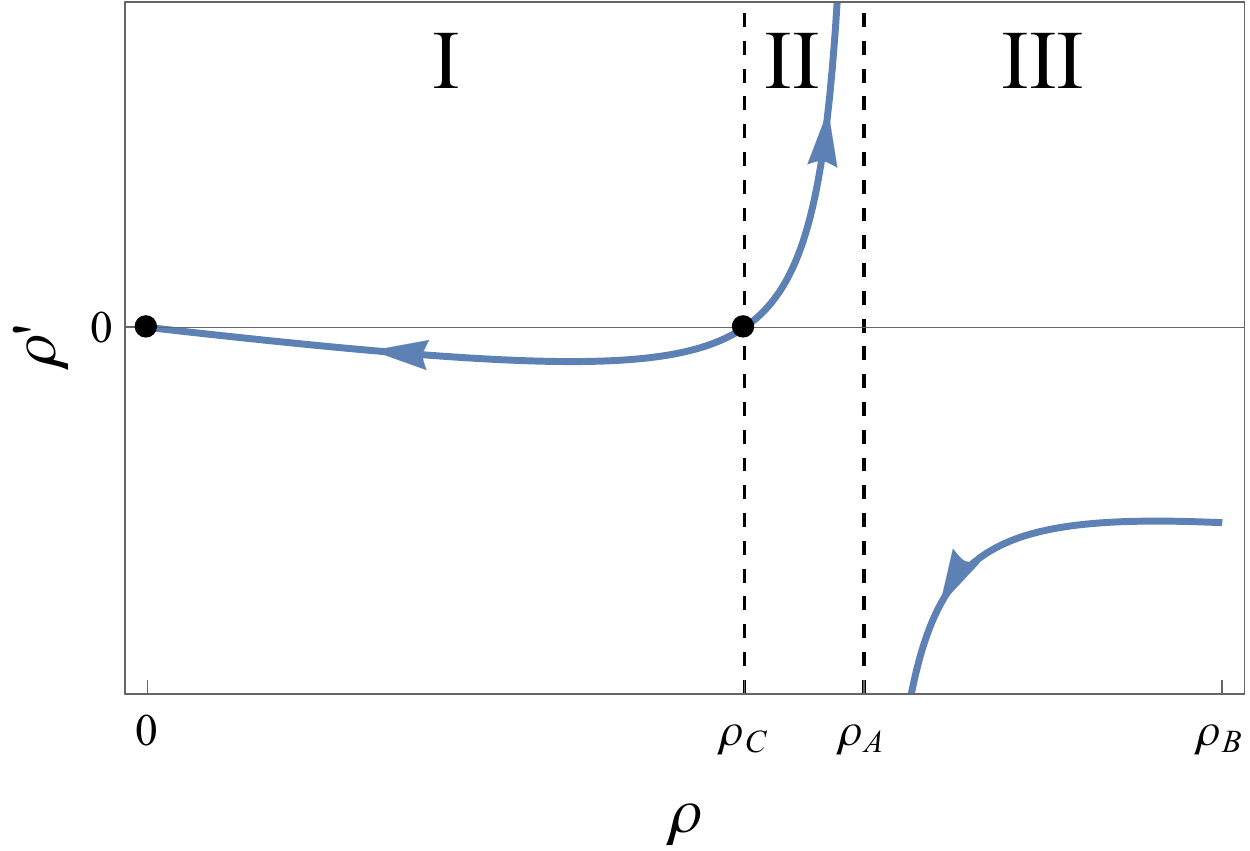} \caption{\label{fig:rho_C < rho_A phase space}A schematic example of the phase
space $(\rho',\rho)$ for the case $\rho_{\text{C}}<\rho_{\text{A}}$.
The dots represent fixed points, while the arrows represent the flow
direction in phase space. The phase space is split into three regions.
Region I corresponds to our Universe, while the bouncing solution
is confined to region III which is disconnected from our Universe.}
\end{figure}

\begin{figure}
\centering{}\includegraphics[width=8.6cm]{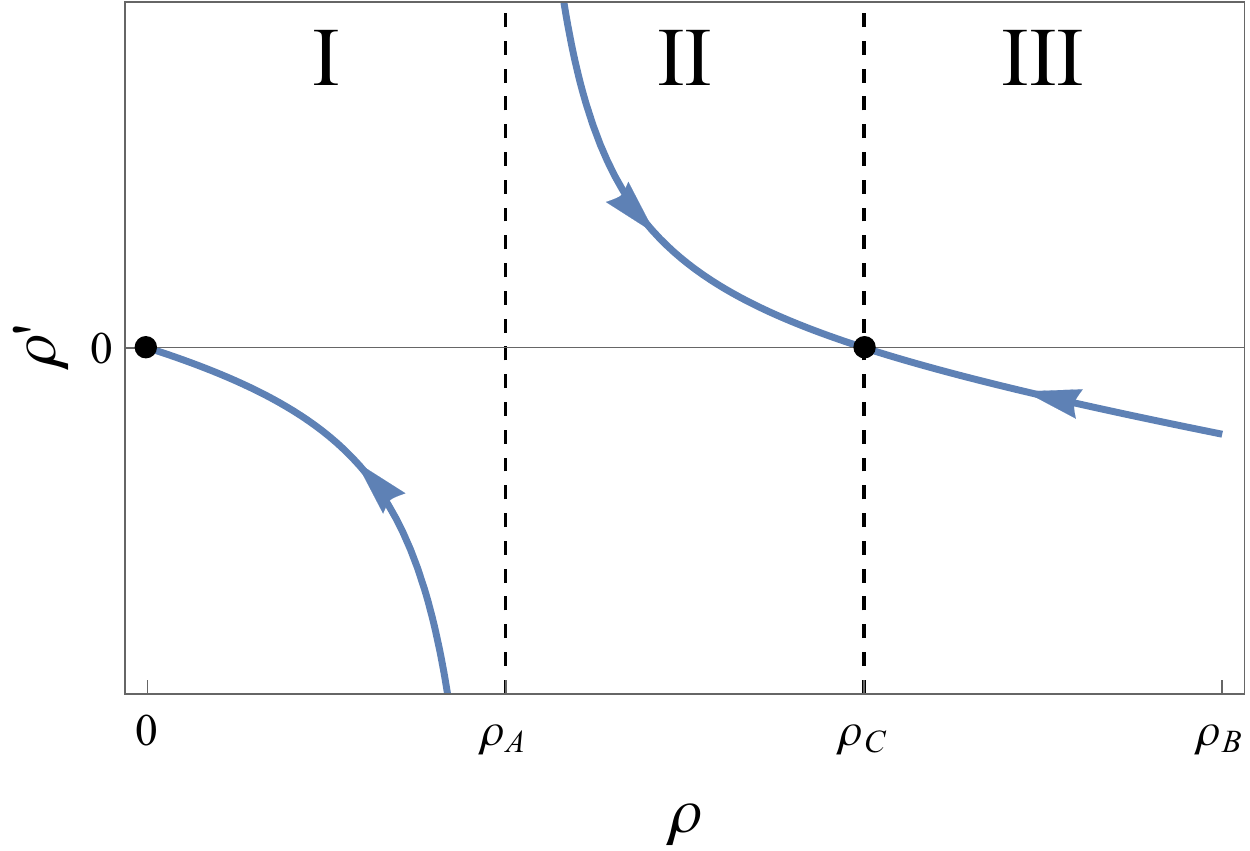}\caption{\label{fig:rho_A < rho_C phase space}A schematic example of the phase
space $(\rho',\rho)$ for the case $\rho_{\text{A}}<\rho_{\text{C}}$.
The dots represent fixed points, while the arrows represent the flow
direction in phase space. The phase space is split into three regions.
Region I corresponds to our Universe, while the bouncing solution
is confined to region III which is disconnected from our Universe.}
\end{figure}

Now our goal is very simple: we want to see if there is a solution
connecting our present density $\rho_{0}$ to the bounce point at
$\rho_{\text{B}}$. For this to happen, we need $\rho_{0}$ and $\rho_{\text{B}}$
to belong to the same phase space region. In other words, we need
the interval $[\rho_{0},\rho_{\text{B}}[$ to be free from fixed or
singular points. From (\ref{eq: rho_C condition}), (\ref{eq: rho_A conditon})
and (\ref{eq: rho prime general}), we see that if $\rho_{\text{A}}\neq\rho_{\text{C}}$,
then we have a fixed point at $\rho_{\text{C}}<\rho_{\text{B}}$ and
also an odd singular point at $\rho_{\text{A}}<\rho_{\text{B}}$.
In this case the bounce at $\rho_{\text{B}}$ is not cosmologically
viable since there are no solutions that connect it to our present
Universe density $\rho_{0}$. Instead, we get either a solution connecting
$\rho_{0}$ to $\rho_{\text{C}}$ if $\rho_{\text{C}}<\rho_{\text{A}}$
as shown in Fig. \ref{fig:rho_C < rho_A phase space}, which takes
an infinite time to reach $\rho_{\text{C}}$ in the past, or a solution
connecting $\rho_{0}$ to $\rho_{\text{A}}$ if $\rho_{\text{A}}<\rho_{\text{C}}$
as shown in Fig. \ref{fig:rho_A < rho_C phase space}, which, similar
to the solution obtained in EMSG, would be past-geodesically incomplete.

The only case remaining now is if $\rho_{\text{A}}=\rho_{\text{C}}$.
We see from (\ref{eq: rho_C condition}), (\ref{eq: rho_A conditon})
and (\ref{eq: rho prime general}) that if $\rho_{\text{A}}=\rho_{\text{C}}$,
then the would-be poles and zeros of $\rho'$ cancel out and $\rho'$
becomes free of any splitting points in the interval $]0,\rho_{\text{B}}]$.
Therefore in this case, $\rho_{\text{B}}$ is a cosmologically viable
bounce.

While our analysis was concerned with bounces, we note for completeness
that the case $\rho_{\text{C}}<\rho_{\text{A}}$, corresponding to
region I in Fig. \ref{fig:rho_C < rho_A phase space}, describes a
viable nonsingular initially de Sitter solution. In this scenario,
the Universe starts and ends with fixed points.

To summarize the results, we have shown that in theories that modify
GR through effective modification of the matter sector, in order to
achieve the usual bounce condition $H=0$ and $\dot{H}>0$ at some
high density $\rho_{\text{B}}$, the theory must have at least one
nontrivial zero for each of $\rho_{\text{eff}}+p_{\text{eff}}$ and
$\frac{d\rho_{\text{eff}}}{d\rho}$ at densities lower than $\rho_{\text{B}}$.
These points would segregate our Universe from the bounce point in
phase space, unless they coincide effectively making $\rho'$ free
of poles and zeros in the interval $]0,\rho_{\text{B}}]$. Therefore,
in addition to the usual bounce conditions of GR, we must have nontrivial
zeros of $\rho_{\text{eff}}+p_{\text{eff}}$ and $\frac{d\rho_{\text{eff}}}{d\rho}$
coincident in the interval $]0,\rho_{\text{B}}]$ to obtain a viable
bounce in these models.

\subsection{Case of $\left(T_{\mu\nu}T^{\mu\nu}\right)^{n}$}

We can now apply the result of our analysis on generalizations of
EMSG that modify the action by adding a term $\left(T_{\mu\nu}T^{\mu\nu}\right)^{n}$
which were studied in \cite{board2017cosmological,EMPG2018}. The
action for these theories (ignoring the cosmological constant) can
be written as 
\begin{equation}
S=\int d^{4}x\,\sqrt{-g}\,\left(\frac{1}{2\kappa}R-\frac{1}{2}\alpha^{2n-1}\,\left(T_{\mu\nu}T^{\mu\nu}\right)^{n}\right)+S_{\text{M}}\,,
\end{equation}
where $\alpha^{-1}$ is the characteristic density scale of this theory,
and we will concern ourselves with $n>\frac{1}{2}$ theories, since
those are the ones that have relevant effects in the high density
regimes.

The effective energy-momentum tensor now becomes 
\begin{equation}
T_{\mu\nu}^{\text{eff}}=T_{\mu\nu}-\frac{1}{2}\alpha^{2n-1}\,\left(T_{\sigma\rho}T^{\sigma\rho}\right)^{n}g_{\mu\nu}+n\alpha^{2n-1}\left(T_{\sigma\rho}T^{\sigma\rho}\right)^{n-1}\Theta_{\mu\nu},
\end{equation}
where $\Theta_{\mu\nu}$ is defined as before in (\ref{eq:Theta mu nu}).
For a perfect fluid with a barotropic equation of state $p=\omega\rho$,
the effective density and pressure become 
\begin{align}
\rho_{\text{eff}} & =\rho-\alpha^{2n-1}\rho^{2n}\left(1+3\omega^{2}\right)^{n-1}\left(\left(n-\frac{1}{2}\right)\left(1+3\omega^{2}\right)+4n\omega\right),\label{eq: rho_eff general n}\\
p_{\text{eff}} & =\rho\omega-\frac{1}{2}\alpha^{2n-1}\rho^{2n}\left(1+3\omega^{2}\right)^{n}.\label{eq: p_eff general n}
\end{align}
We can easily see that $\rho_{\text{eff}}$ satisfies our assumptions
(\ref{eq: bounce cond 1}) and (\ref{eq: low density cond 1}) for
$\omega\geq0$ (which is the case we care about for the moment at
least), and it would have a profile similar to the one in Fig. \ref{fig:rhp_eff profile example}.
From (\ref{eq: rho_eff general n}) and (\ref{eq: p_eff general n})
we get 
\begin{equation}
\rho_{\text{eff}}+p_{\text{eff}}=\rho(1+\omega)\left(1-n(3\omega+1)\alpha^{2n-1}\rho^{2n-1}\left(3\omega^{2}+1\right)^{n-1}\right),\label{eq: rho+p eff general n}
\end{equation}
\begin{equation}
\frac{d\rho_{\text{eff}}}{d\rho}=1-2n\alpha^{2n-1}\rho^{2n-1}\left(1+3\omega^{2}\right)^{n-1}\left(\left(n-\frac{1}{2}\right)\left(1+3\omega^{2}\right)+4n\omega\right).\label{eq: d rho_eff by d rho general n}
\end{equation}
Note that adding a cosmological constant $\Lambda$ would not change
the latter expressions. Adding $\Lambda$ is equivalent to the following
transformation 
\begin{align}
\rho_{\text{eff}} & \rightarrow\rho_{\text{eff}}+\frac{\Lambda}{\kappa},\nonumber \\
p_{\text{eff}} & \rightarrow p_{\text{eff}}-\frac{\Lambda}{\kappa},
\end{align}
and $\rho_{\text{eff}}+p_{\text{eff}}$ and $\frac{d\rho_{\text{eff}}}{d\rho}$
are clearly invariant under such a transformation.

By comparing the two expressions in (\ref{eq: rho+p eff general n})
and (\ref{eq: d rho_eff by d rho general n}), in order to get the
nontrivial zero of $\rho_{\text{eff}}+p_{\text{eff}}$ to coincide
with the zero of $\frac{d\rho_{\text{eff}}}{d\rho}$, we see that
the only nontrivial value for $n$ that satisfies that condition is
(assuming $\omega\geq0$) 
\begin{equation}
n=\frac{3\omega^{2}+3\omega+2}{6\omega^{2}+8\omega+2}.\label{eq: n that gives bounce}
\end{equation}
This value for $n$ leaves the conservation equation unmodified from
the one we know in GR (for a single-component fluid with this particular
value of $\omega$), namely 
\begin{equation}
\rho^{'}+3\rho\left(1+\omega\right)=0.
\end{equation}
For radiation ($\omega=\frac{1}{3}$) that value is $n=\frac{5}{8}$,
and hence, $\left(T_{\mu\nu}T^{\mu\nu}\right)^{5/8}$ is the only
theory in this class that gives a viable (radiation-dominated) bounce.
By solving for the nontrivial zero of $\rho_{\text{eff}}$ in (\ref{eq: rho_eff general n})
with $n=\frac{5}{8}$ and $\omega=\frac{1}{3}$, the density at the
bounce in the $n=\frac{5}{8}$ theory will be equal to $\frac{8}{3\sqrt{3}\alpha}$.
Further analysis of this model may be required to ensure that it can
reproduce other aspects of standard cosmology; for example, it is
important to check the stability of this solution against inhomogeneous
perturbations as it may lead to some instabilities similar to those
discussed in \cite{peter2001}. In particular, we note that while
the $n=5/8$ model avoids any singular behavior in the continuity
equation, it is very likely that the effective Euler equation will
have singular points due to the nonlinearity of $p_{\text{eff}}$
in $\rho$ and $p$.

It is interesting to note from (\ref{eq: n that gives bounce}) that
for EMSG ($n=1$), we can have a bounce for a dust-only ($\omega=0$)
universe.

Finally, we note for completeness that the case $1/2<n<5/8$ is the
case corresponding to region I in Fig. \ref{fig:rho_C < rho_A phase space}
which provides a geodesically complete nonsingular solution that can
describe our Universe. In these theories the initial singularity is
replaced by a de Sitter fixed point (at $t\rightarrow-\infty$), as
a result, the Universe is going to interpolate between two fixed points
one at a high density and another at a low or vanishing density. It
would be interesting to study such theories in future works, particularly
with the interpretation of $\alpha^{-1}$as the Planck scale in that
case.

\section{Junction Conditions in EMSG\label{sec:Junction-Conditions}}

As we recall, solving the conservation equation (\ref{eq:conseq radiation only})
and the Friedmann equation (\ref{eq:a_dot ODE}) led to the following
branches of solutions for the scale factor 
\begin{equation}
a(t)=a_{\text{min}}\sqrt{1\pm2H_{\text{max}}t},\,\,\,\,\,\,\,\,\pm t\geq0,\label{eq: a(t) branches}
\end{equation}
and the following (independent) branches for $\rho_{r}$: 
\begin{equation}
\rho_{r}(a)=\frac{1}{4\alpha}\left(1\pm\sqrt{1-\left(\frac{a_{\text{min}}}{a}\right)^{4}}\right).\label{eq:rho_r branches}
\end{equation}
In Sec. \ref{sec:Cosmology-in-EMSG}, we picked the positive branch
for $a$ to get a solution valid for $t>0$, and we picked the negative
branch for $\rho_{r}$ to get a solution that corresponds to $\rho_{r}\rightarrow0$
as $a\rightarrow\infty$. This led to a geometrically nonsingular
solution, albeit geodesically incomplete. In this section we will
join that solution with the other branch using appropriate junction
conditions in order to get a geodesically complete solution, albeit
with a curvature singularity at $t=0$. Since EMSG inherits the geometric
side of GR, standard junction conditions in GR will be utilized in
this section; the reader can be referred to \cite{poisson2004relativist}
for a review.

Using the FRW coordinates, we can define a spacelike hypersurface
$\Sigma$ at $t=0$. This hypersurface now splits the spacetime into
two regions with $t>0$ and $t<0$ respectively. We can define the
following useful notation for the jump in a tensor $A$ across $\Sigma$
\begin{equation}
[A]\equiv A(0^{+})-A(0^{-}).
\end{equation}
Here, we have two solutions, one in the region where $t>0$ and the
other is in the region where $t<0$; we would like to join them at
$\Sigma$. In order to achieve this smoothly, we need two conditions.
The first junction condition is 
\begin{equation}
[g_{\mu\nu}]=0.\label{eq: 1st jun condition}
\end{equation}
The continuity of the metric here is a very important condition as
otherwise the Christoffel symbols would have Dirac deltas, and the
curvature tensors then would be ill defined. The second junction condition
is 
\begin{equation}
[K_{\mu\nu}]=0,
\end{equation}
where $K_{\mu\nu}$ is the extrinsic curvature. This condition is
necessary for a smooth transition across $\Sigma$; however, a finite
jump in $K_{\mu\nu}$ is sufficient for geodesic extension, but it
will cause a curvature singularity at $\Sigma$. This singularity
has the physical interpretation of having a surface energy momentum
tensor at $\Sigma$, which is given by 
\begin{equation}
S_{\mu\nu}=\frac{1}{\kappa}\left([K_{\mu\nu}]-[K]h_{\mu\nu}\right),\label{eq: surface EM tensor}
\end{equation}
where $K$ is the trace of the extrinsic curvature and $h_{\mu\nu}$
is the induced metric on $\Sigma$.

We can now start joining the two solutions in (\ref{eq: a(t) branches})
at $\Sigma$, we simply get 
\begin{equation}
a(t)=a_{\text{min}}\sqrt{1+2H_{\text{max}}|t|}.\label{eq: a(t) extended solution}
\end{equation}
We can see that this automatically satisfies the continuity of the
metric condition (\ref{eq: 1st jun condition}). The Hubble rate then
becomes 
\begin{equation}
H(t)=\frac{H_{\max}}{1+2H_{\text{max}}|t|}\text{sgn}(t).
\end{equation}
We can see that the Hubble rate has a finite jump at $t=0$. For the
FRW metric with our choice of $\Sigma$, we only need to focus on
the spatial components of the extrinsic curvature and the induced
metric, which are given by 
\begin{equation}
h_{ij}=g_{ij}=a(t)^{2}\delta_{ij},
\end{equation}
\begin{equation}
K_{ij}=\frac{1}{2}\partial_{t}h_{ij}=a(t)^{2}H(t)\delta_{ij}.
\end{equation}
We can see that the extrinsic curvature picks a finite jump from $H$,
which we can calculate as follows 
\begin{equation}
[K_{ij}]=2a_{\text{min}}^{2}H_{\text{max}}\delta_{ij},
\end{equation}
\begin{equation}
[K]=6H_{\text{max}}.
\end{equation}

Since we have a finite jump in the extrinsic curvature, we get a surface
energy-momentum tensor contribution (\ref{eq: surface EM tensor})
as 
\begin{equation}
S_{ij}=-\frac{4}{\kappa}a_{\text{min}}^{2}H_{\text{max}}\delta_{ij}.\label{eq: Sij expression}
\end{equation}
In GR, this surface energy-momentum tensor would be a contribution
to the ordinary energy momentum tensor; however, in the case of EMSG,
since the Einstein tensor is sourced by the effective energy momentum
tensor instead, (\ref{eq: Sij expression}) is a contribution to the
effective energy-momentum tensor, i.e. we have a term like 
\begin{equation}
T_{ij}^{\text{eff}}\Bigr|_{\Sigma}=\delta(t)S_{ij}.
\end{equation}
This is a surface pressure term that is added to the normally occurring
$p_{\text{eff}}$ mentioned before. Therefore, the total effective
pressure is singular at $\Sigma$. Since the effective pressure is
singular at $\Sigma$ while the effective density is finite (which
can simply be shown from the Friedmann equation), the singularity
at $\Sigma$ is a sudden singularity \cite{Barrow2004_sudden}. This
type of singularities is known to be weak (and hence geodesically
extendible) according to Tipler and Krolak's definitions \cite{Tipler_1977,Krolak_1985}
as was shown in \cite{Jambarina2004_sudden}. Furthermore, the geodesic
extendibility here would be the same as in the case considered in
\cite{Awad2016_Weyl} since $a(t)$ in both cases have the same Puiseux
expansion up to first order in $t$ (for a more detailed account on
the behavior of geodesics according to the Puiseux expansion of the
scale factor, see \cite{jambarina2007geodesic}).

Finally, the solution for the density in the region where $t<0$ can
be either branch in (\ref{eq:rho_r branches}). So the extended solution
for all $t$ can either be 
\begin{equation}
\rho_{r}(t)=\frac{1}{4\alpha}\left(1-\sqrt{1-\left(\frac{a_{\text{min}}}{a(t)}\right)^{4}}\right),
\end{equation}
or 
\begin{equation}
\rho_{r}(t)=\frac{1}{4\alpha}\left(1-\text{sgn}(t)\,\sqrt{1-\left(\frac{a_{\text{min}}}{a(t)}\right)^{4}}\right),
\end{equation}
where $a(t)$ is given by (\ref{eq: a(t) extended solution}).

It is important to note that the junction conditions in this analysis
depended only on two features of the solution; namely, $a(0)=a_{\text{min}}$
and $H(0^{\pm})=\pm H_{\text{max}}$, rather than the full behavior
of $a(t)$. These two features are also in the $\left(T_{\mu\nu}T^{\mu\nu}\right)^{n}$
theories with $n>\frac{5}{8}$, and thus they would have the same
junction conditions as in this analysis in terms of $a_{\text{min}}$
and $H_{\text{max}}$; the expressions for the latter parameters depend
on the choice of $n$ of course.

\section{Conclusion\label{sec:Conclusion}}

EMSG was first proposed as a theory that cures the initial cosmological
singularity, reminiscent of the behavior of theories like loop quantum
gravity. We have shown in this work that the regular-bouncing solution
one can obtain in such a theory is not viable for our Universe. Instead,
the viable solution branch, while having no curvature singularities,
is only valid up to a certain point in the past. This branch can be
joined with its contracting counterpart using the junction conditions
outlined in Sec. \ref{sec:Junction-Conditions} to get a fully extended
solution; however, the only way to achieve such an extension is by
having a (weak) singularity at the junction. In light of this solution,
we see that EMSG can at best provide a singular-bouncing solution,
and thus the similarity to theories like loop quantum gravity is only
superficial.

The singularity in the extended solution---or the geodesic incompleteness
in the non-extended one---suggests that EMSG needs to be corrected
at density scales close to $\alpha^{-1}$. This means that EMSG should
be interpreted as an effective field theory, valid only at scales
away from $\alpha^{-1}$, and one expects new (gravitational) physics
to appear at scales at (or beyond) $\alpha^{-1}$. While these new
physics do not necessarily have to be quantum, it is more natural
to assume that new gravitational physics arise at the Planck scale,
and this motivates the interpretation of $\alpha^{-1}$ as the Planck
density.

We have also seen that theories that modify GR by effectively modifying
the matter Lagrangian must satisfy the stringent condition outlined
in Sec. \ref{sec:Bounce-Analysis-for General theories} in order to
have a viable regular-bouncing solution. For the case of $\left(T_{\mu\nu}T^{\mu\nu}\right)^{n}$
generalizations of EMSG, only the $n=5/8$ case satisfies that condition.
Aside from bounces, we have shown that theories with $1/2<n<5/8$
can provide a viable nonsingular initially de Sitter solution. It
would be interesting to construct arguments similar to those in Sec.
\ref{sec:Bounce-Analysis-for General theories} for more general theories
that have a total Lagrangian of the form $f(R,T_{\mu\nu}T^{\mu\nu})$
which would have a more complicated phase space structure.

While only studying the cosmological aspects of EMSG, we have encountered
singular points in the matter differential equations due to the nonlinearities
introduced in the theory; similar singular behavior can occur in any
other physical situation. These singular points can split the phase
space, similar to what happened in the cosmology of EMSG, which can
cause geodesic incompleteness. Therefore, even for theories that satisfy
the condition in Sec. \ref{sec:Bounce-Analysis-for General theories},
which was obtained for the case of an isotropic universe with perfect
fluid content, they may not be valid for all physical scenarios at
scales close to their characteristic density scale ($\alpha^{-1}$
in the case of EMSG and its generalizations). 
\begin{acknowledgments}
Some expressions in this work were computed or checked using the ``xAct''
Mathematica packages \cite{xact_martin2004}. A. H. Barbar was supported
in part by the American University in Cairo, Research Grant Agreement
807 No. SSE-PHYS-M.AFY18- RG(2-17)-2017-Feb-12-21- 808 41-43. 
\end{acknowledgments}

\appendix

\section{Effective Energy-Momentum Tensor\label{sec:Appendix_Effective-Energy-Momentum-Tensor}}

The ordinary energy momentum tensor is defined as 
\begin{equation}
T_{\mu\nu}=-\frac{2}{\sqrt{-g}}\frac{\delta\left(\sqrt{-g}L_{m}\right)}{\delta g^{\mu\nu}},
\end{equation}
which gives 
\begin{equation}
T_{\mu\nu}=L_{m}g_{\mu\nu}-2\frac{\partial L_{m}}{\partial g^{\mu\nu}},
\end{equation}
and its variation with respect to the metric is 
\begin{align}
\frac{\delta T_{\sigma\rho}}{\delta g^{\mu\nu}} & =-g_{\sigma\mu}g_{\rho\nu}L_{m}+g_{\sigma\rho}\frac{\partial L_{m}}{\partial g^{\mu\nu}}-2\frac{\partial^{2}L_{m}}{\partial g^{\mu\nu}\partial g^{\sigma\rho}}\\
 & =-g_{\sigma\mu}g_{\rho\nu}L_{m}+g_{\sigma\rho}\left(\frac{1}{2}L_{m}g_{\mu\nu}-\frac{1}{2}T_{\mu\nu}\right)-2\frac{\partial^{2}L_{m}}{\partial g^{\mu\nu}\partial g^{\sigma\rho}}\\
 & =\left(\frac{1}{2}g_{\sigma\rho}g_{\mu\nu}-g_{\sigma\mu}g_{\rho\nu}\right)L_{m}-\frac{1}{2}T_{\mu\nu}g_{\sigma\rho}-2\frac{\partial^{2}L_{m}}{\partial g^{\mu\nu}\partial g^{\sigma\rho}}.\label{eq:var of T wrt g_mu_nu}
\end{align}
If we have a theory with the following total action 
\begin{equation}
S=\int d^{4}x\,\sqrt{-g}\,\left(\frac{1}{2\kappa}R+L_{m}+F(T_{\sigma\rho}T^{\sigma\rho})\right),
\end{equation}
it will be equivalent to having an effective matter Lagrangian as
\begin{equation}
L_{m,\text{eff}}=L_{m}+F(T_{\sigma\rho}T^{\sigma\rho}).
\end{equation}
Thus, the effective energy momentum tensor will be 
\begin{align}
T_{\mu\nu}^{\text{eff}} & \coloneqq-\frac{2}{\sqrt{-g}}\frac{\delta\left(\sqrt{-g}\,L_{m,\text{eff}}\right)}{\delta g^{\mu\nu}}\\
 & =T_{\mu\nu}-\frac{2}{\sqrt{-g}}\frac{\delta\left(\sqrt{-g}\,F(T_{\sigma\rho}T^{\sigma\rho})\right)}{\delta g^{\mu\nu}}\\
 & =T_{\mu\nu}+F(T_{\sigma\rho}T^{\sigma\rho})g_{\mu\nu}-2F_{\boldsymbol{T^{2}}}\Theta_{\mu\nu},
\end{align}
where 
\begin{equation}
F_{\boldsymbol{T^{2}}}\equiv\frac{\partial F}{\partial(T_{\sigma\rho}T^{\sigma\rho})},
\end{equation}
and 
\begin{equation}
\Theta_{\mu\nu}\equiv\frac{\delta\left(T_{\sigma\rho}T^{\sigma\rho}\right)}{\delta g^{\mu\nu}}.
\end{equation}
Now what remains is to calculate $\Theta_{\mu\nu}$ as follows 
\begin{align}
\Theta_{\mu\nu} & =2T_{\mu\sigma}T_{\,\,\nu}^{\sigma}+2T^{\sigma\rho}\frac{\delta T_{\sigma\rho}}{\delta g^{\mu\nu}}\\
 & =2T_{\mu\sigma}T_{\,\,\nu}^{\sigma}+2T^{\sigma\rho}\left(\left(\frac{1}{2}g_{\sigma\rho}g_{\mu\nu}-g_{\sigma\mu}g_{\rho\nu}\right)L_{m}-\frac{1}{2}T_{\mu\nu}g_{\sigma\rho}-2\frac{\partial^{2}L_{m}}{\partial g^{\mu\nu}\partial g^{\sigma\rho}}\right)\\
 & =2T_{\mu\sigma}T_{\,\,\nu}^{\sigma}-2L_{m}\left(T_{\mu\nu}-\frac{1}{2}Tg_{\mu\nu}\right)-TT_{\mu\nu}-4T^{\sigma\rho}\frac{\partial^{2}L_{m}}{\partial g^{\mu\nu}\partial g^{\sigma\rho}},
\end{align}
where we have used the result of (\ref{eq:var of T wrt g_mu_nu})
in the second line.

\section{Matter-Radiation Fluid in EMSG\label{sec:Appendix_Two-component-Fluid-in}}

In EMSG, if we consider a matter-radiation perfect fluid in FRW spacetime,
we can get the individual conservation equation for each component
by applying the results in (\ref{eq: cons eq fluid 1}) on radiation
($\omega=1/3$) and matter ($\omega=0$) respectively, thus we get

\begin{equation}
\left(1-4\alpha\rho_{r}-\frac{7\alpha\rho_{m}}{3}\right)\rho_{r}'+4\rho_{r}-8\alpha\rho_{r}^{2}-5\alpha\rho_{m}\rho_{r}=0,\label{ceqr}
\end{equation}
\begin{equation}
\left(1-\alpha\rho_{m}-\frac{7\alpha\rho_{r}}{3}\right)\rho_{m}'+3\rho_{m}-3\alpha\rho_{m}^{2}-5\alpha\rho_{r}\rho_{m}=0.\label{ceqm}
\end{equation}

\subsubsection{Matter domination}

In the case of matter domination, the matter conservation equation
(\ref{ceqm}) becomes 
\begin{align}
\left(1-\alpha\rho_{m}\right)\rho_{m}'+3\rho_{m}-3\alpha\rho_{m}^{2} & =0\\
\Rightarrow\left(1-\alpha\rho_{m}\right)\left(\rho_{m}'+3\rho_{m}\right) & =0.
\end{align}
This gives the same behavior as GR, namely $\rho_{m}(a)=\rho_{m0}\,a^{-3}$,
where $\rho_{m0}$ is the present matter density and the present scale
factor has been set to unity.

The radiation conservation equation (\ref{ceqr}) becomes 
\begin{equation}
\left(1-\frac{7\alpha\rho_{m}}{3}\right)\rho_{r}'+4\rho_{r}-5\alpha\rho_{m}\rho_{r}=0,
\end{equation}
which gives the solution 
\begin{equation}
\rho_{r}(a)=\rho_{r0}\,a^{-4}\left(\frac{1-\frac{7}{3}\alpha\rho_{m0}}{1-\frac{7}{3}\alpha\rho_{m0}a^{-3}}\right)^{13/21}.
\end{equation}
In order for $\rho_{r}$ to be real valued, we must have $a\geq\left(\frac{7}{3}\alpha\rho_{m0}\right)^{1/3}$.
Therefore we must have a minimum scale in this scenario, which is
$a=\left(\frac{7}{3}\alpha\rho_{m0}\right)^{1/3}$. Notice that $\rho_{r}$
diverges as $a\rightarrow\left(\frac{7}{3}\alpha\rho_{m0}\right)^{1/3}$,
while $\rho_{m}$ remains finite. Thus, as we get closer to this minimum
scale, the radiation part dominates, which contradicts our assumption
that we are working in the regime of matter domination. Therefore
we can conclude that this solution, which corresponds to a matter
dominating era, can only be valid at scales much larger than $\left(\frac{7}{3}\alpha\rho_{m0}\right)^{1/3}$.
This gives us a hint that matter dominates away from the early Universe
in this theory.

\subsubsection{Radiation domination}

In the case of radiation domination, the conservation equations for
radiation (\ref{ceqr}) and matter (\ref{ceqm}) respectively become
\begin{equation}
\left(1-4\alpha\rho_{r}\right)\rho_{r}'+4\rho_{r}-8\alpha\rho_{r}^{2}=0,\label{eq:rad dom rad}
\end{equation}
\begin{equation}
\left(1-\frac{7\alpha\rho_{r}}{3}\right)\rho_{m}'+3\rho_{m}-5\alpha\rho_{r}\rho_{m}=0.
\end{equation}
These lead to solutions 
\begin{equation}
\rho_{m}=\rho_{m0}\left(\frac{3-7\alpha\rho_{r}}{3-7\alpha\rho_{r0}}\right)^{15/14}\left(\frac{1-2\alpha\rho_{r0}}{1-2\alpha\rho_{r}}\right)^{3/4}\left(\frac{\rho_{r}}{\rho_{r0}}\right)^{3/4}\label{eq:matter in radiation dom}
\end{equation}
\begin{equation}
\rho_{r}=\frac{1}{4\alpha}\left(1-\sqrt{1-\left(\frac{a_{\text{min}}}{a}\right)^{4}}\right),
\end{equation}
where $a_{\text{min}}\equiv\left(8\alpha\rho_{r0}\left(1-2\alpha\rho_{r0}\right)\right)^{1/4}$,
and again we have solved using the present values for matter $\rho_{m0}$
and radiation $\rho_{r0}$ while setting the present scale factor
$a_{0}$ to unity. We note that the use of present values as conditions
in this approximation is still justified since EMSG is expected to
coincide with GR at some point in the early Universe; after that point
the original equations (\ref{ceqr}) and (\ref{ceqm}) will decouple
anyway and reduce to their GR counterparts, and hence any condition
valid for solving the GR equations after that point is also valid
for EMSG, regardless of what component dominates at that condition.

We see from (\ref{eq:matter in radiation dom}) that even at high
radiation densities ($\rho_{r}\sim\frac{1}{4\alpha}$), we have 
\begin{equation}
\rho_{m}\sim\rho_{m0}\left(\frac{\rho_{r}}{\rho_{r0}}\right)^{3/4},\label{eq:rho_m relation with rho_r}
\end{equation}
which is the same behavior of matter as in GR. Therefore the mere
requirement that EMSG coincides with GR before the end of the standard
radiation dominated era, which is required in order for EMSG to be
cosmologically viable, is sufficient for having early-time radiation
domination in EMSG. For example, the consistency of EMSG with GR was
checked in \cite{NeutronStarstConstraintEMSG2018} where they constrained
the parameter $\alpha$ from neutron star observations; their constraint
translates to our definition for $\alpha$ (which differs from theirs
by a factor of $-1/2$) as 
\begin{equation}
0\leq\alpha\lesssim10^{-38}\,\text{erg}^{-1}\,\text{cm}^{3},\label{eq: alpha constraint NS}
\end{equation}
where we only quoted the $\alpha\geq0$ part of the constraint as
it is the one relevant in our case. They then showed that under the
upper bound of this constraint, radiation cosmology in EMSG is consistent
with standard cosmology up to energy-density scales as high as $10^{-34}\,\text{erg}\,\text{cm}^{-3}$
and time scales as early as $10^{-4}\,\text{s}$. Given that, our
result in (\ref{eq:matter in radiation dom}) shows that matter density
values would be the same as those in GR (corrections would be extremely
small due to the tight constraint on $\alpha$). This automatically
means that radiation dominates in the early Universe in EMSG as it
does in GR.

 \bibliographystyle{apsrev4-1}
\bibliography{viability}

\end{document}